\documentclass[epj]{svjour}
\usepackage{graphicx,graphics}
\usepackage{amsmath,amssymb,amsfonts}
\usepackage{color}
\usepackage{epsfig}
\usepackage{latexsym}
\usepackage{float}
\usepackage{subfig}

\definecolor{linkcolor}{rgb}{0,0,0.66} %hyperlink
\usepackage[pdftex,colorlinks=true, pdfstartview=FitV, linkcolor=linkcolor, citecolor=linkcolor, urlcolor=linkcolor, hyperindex=true,hyperfigures=true]{hyperref} %hyperlink

\newcommand{\caO}{{\mathcal O}}

\renewcommand{\Pr}{{\mathcal P}}

\def\beq{\begin{equation}}
\def\eeq{\end{equation}}
\def\bea{\begin{eqnarray}}
\def\eea{\end{eqnarray}}
\def\ba{\begin{array}}
\def\ea{\end{array}}

\begin{document}
\title{Thermal response of a Fermi-Pasta-Ulam chain with Andersen thermostats}
\author{Federico D'Ambrosio\inst{1} \and Marco Baiesi\inst{2,3} } 
    
\institute{Department of Information and Computing Science, Universiteit Utrecht, Princetonplein 5, 3584 CC Utrecht, The Netherlands
\and
Department of Physics and Astronomy, University of Padova, Via Marzolo 8, I-35131 Padova, Italy
 \and 
INFN, Sezione di Padova, Via Marzolo 8, I-35131 Padova, Italy}

\authorrunning{F. D'Ambrosio and M. Baiesi}
\titlerunning{Thermal response of a Fermi-Pasta-Ulam chain with Andersen thermostats}
\abstract{
The linear response to temperature variations is well characterised for equilibrium systems but a similar theory is not available, for example, for inertial heat conducting systems, whose paradigm is the Fermi-Pasta-Ulam (FPU) model driven by two different boundary temperatures. For models of inertial systems out of equilibrium, including relaxing systems, we show that Andersen thermostats are a natural tool for studying the thermal response. We derive a fluctuation-response relation that allows to predict thermal expansion coefficients or the heat capacitance in nonequilibrium regimes. Simulations of the FPU chain of oscillators suggest that estimates of susceptibilities obtained with our relation are better than those obtained via a small perturbation.
 % end of PACS codes
} %end of abstract
\maketitle

\section{Introduction}

The thermal response is usually associated with coefficients such as the specific heat or the thermal expansion coefficient. In equilibrium these coefficients may be computed by means of fluctuation-response formulas, in which the response to temperature variations is related to the natural fluctuations of the systems. For instance, the specific heat is the variance of the energy in equilibrium. As encoded in the Kubo formula~\cite{tod92}, the response of equilibrium systems may be entirely described by its dissipation, or entropy production~\cite{bai13}, in the transient toward a new equilibrium after the perturbation.

The thermal response of systems out of equilibrium is less understood. For example, the equilibrium theory cannot be used to fully determine how glasses undergoing a relaxation do respond to an increase of temperature~\cite{bar99,may06,mag10}. 
Another instance of nonequilibrium regime is a system carrying a heat flow from a hot to a cold reservoir, say a cantilever heated by a laser at one end~\cite{gei17} or larger oscillating devices used in gravitational wave detectors~\cite{con13}. A change in the laser intensity, translated into a change of one boundary temperature, would lead to forms of thermal response that, again, cannot be described within the framework of equilibrium systems. 

Studies on nonequilibrium linear response~\cite{cug94,nak08,che08z,vil09,spe09,sei10,pro09,ver11a,lip07,alt16,bai09,bai09b,bai10} were recently complemented by specific works on thermal response~\cite{che09b,bok11,bra15,pro16,bai14,yol16,fal16,fal16b,bai16}. In particular, a linear response approach first developed for the case of mechanical perturbations~\cite{bai09,bai09b,bai10} was recently extended to include the response to temperature variations~\cite{bai14,yol16,fal16,fal16b,bai16,yol17}. Those works dealt with overdamped stochastic systems and provided fluctuation-response relations based on standard integrals, thus regularizing singular terms present in the earlier contribution~\cite{bai14}. However, currently we do not have similar results for inertial systems.
This means that the basic models of heat conduction, such as the classical driven Fermi-Pasta-Ulam (FPU) chain of coupled oscillators~\cite{lep03,dha08}, cannot be described with a linear response theory for temperature variations. In fact, to our knowledge, despite the extensive amount of studies focusing on FPU models, there is no theory describing their thermal response to a change of one of the two boundary temperatures.

We derive a thermal fluctuation-response relation for (nonequilibrium) inertial systems driven by Andersen thermostats~\cite{and80,frenkel}. These thermostats are a standard tool for equilibrium molecular dynamics simulations in the canonical ensemble. We show that, as a key advantage, the Andersen thermostats do not bring all mathematical problems of Langevin heat baths, yet they bring to expressions displaying the physically relevant quantities, such as the entropy production in the reservoirs. The fluctuation-response relation we obtain is based on the stochastic nature of Andersen thermostats, which allows us to apply a well-established scheme for determining response functions and susceptibilities from ratios of path-weights. As a numerical example, we show two forms of susceptibility of the FPU system to a variation of one boundary temperature.

\section{Thermal susceptibility of systems driven by Andersen thermostats}
\label{sec:susceptibility}

We consider systems evolving via Newton's equations
\begin{align}
\dot x_i &= v_i \nonumber \\
m_i \dot v_i &= F_i(\{x_i\})
\end{align}
where $x_i$, $v_i$, $m_i$, and $F_i$ are, respectively, positions, velocities, masses and forces.

A thermalization of a degree of freedom $i$ to a temperature $T_i$ is achieved by a Andersen thermostat \cite{and80,frenkel}.
The velocity $v_i$ is updated at some instants with a value $v$ extracted from an equilibrium one-dimensional Maxwell-Boltzmann distribution
\begin{equation}
\rho_{eq} (v) = \sqrt{\frac{m_i}{2 \pi T_i}} e^{-\frac{m v^2}{2 T_i}}
\label{eq:MB}
\end{equation}
where the Boltzmann constant is set $k_B=1$.
As in Markovian jump processes, instants of updates are separated by time intervals $\Delta t$ extracted from an exponential distribution
\begin{equation}
P(\Delta t) \propto e^{- \frac{\Delta t}{\tau}}
\label{eq:PDT}
\end{equation}
where $\tau$ is a parameter of the simulation.

We first study the response to the perturbation of a single temperature 
\[
T_i \rightarrow T_i + \theta \,.
\]
activated at time $0$ and persistent till time $t$.
A simultaneous perturbation of several temperatures would be simply be a linear combination of the following basic results, as shown at the end of this section.
A trajectory $\omega=\{x_i(s),v_i(s)\}$, $s\in[0,t]$ has a path-weight denoted by $\Pr[\omega]$ and related averages are written as $\langle \ldots \rangle$. The starting point of a trajectory is from an arbitrary initial distribution $\rho(\{x_i(0),v_i(0)\})$ 
which is not recalled explicitly in the notation (in general it can also be non-steady state distribution).
The notation for perturbed trajectories, which start from the same $\rho$ but evolve with modified $T_i + \theta$, is turned to $\Pr_\theta[\omega]$ and $\langle \ldots \rangle_\theta$.

For an observable $\caO(t)$ by definition we have a thermal susceptibility 
\begin{equation}
\label{eq:direct-chi}
\chi_\caO (t) 
\equiv 
\left.
\frac{\partial}{\partial \theta}\langle \caO(t)\rangle_\theta
\right|_{\theta=0}
\end{equation}
While $\caO$ could be a functional $\caO[\omega]$ of the whole trajectory, we keep the notation $\caO(t)$ to highlight the evaluation of such observable after a time $t$ from the formal  activation of the perturbation $\theta$.
Let us express the mean perturbed value of $\caO$ by
\begin{equation}
\label{eq:meanO}
\langle \caO(t)\rangle_\theta = \sum_{\omega} \Pr_\theta[\omega]  \caO(t) \,,
\end{equation}
where $\sum_{\omega}$ is a simplified notation for indicating a formal inclusion in the statistics of all (uncountable) trajectories of the system.
To write this susceptibility as a function of unperturbed correlation functions, we rewrite (\ref{eq:meanO}) as
\begin{equation}
\label{eq:meanO2}
\langle \caO(t)\rangle_\theta
 = \sum_{\omega} \Pr [\omega] \frac{\Pr_\theta}{\Pr}[\omega] \caO(t)
\end{equation}
which is an average in the unperturbed system, namely
\begin{equation}
\langle \caO(t)\rangle_\theta
= \left\langle \frac{\Pr_\theta}{\Pr}[\omega] \caO(t) \right\rangle
\end{equation}
To compute the path weights ratio, note that the velocity updates are stochastic only at instants when they are replaced by values extracted from Maxwell-Boltzmann distributions; the deterministic evolution in between these updates is the same in the perturbed and unperturbed dynamics. Thus the path-weight ratio $\Pr_\theta/\Pr$ can be written as a product over all velocity jumps of the velocity probabilities (perturbed and unperturbed) at every jump, which gives factors different from $1$ only for the degree of freedom $i$ driven by the perturbed reservoir.
In a trajectory $\omega$ in a time span $t$ with $n_i$ updates of $v_i$, we use the index $\alpha$ for velocity updates and denote by $v_i^{\alpha+}$ the velocity right after the jump $\alpha$, that is the velocity extracted from the equilibrium distribution in the Andersen scheme. With this notation we have
\begin{align}
\frac{\Pr_\theta}{\Pr} [\omega] 
&= 
\prod_{\alpha}^{n_i} 
\frac{\sqrt{\frac{m}{2 \pi (T_i + \theta)}}e^{-\frac{m_i (v_i^{\alpha+})^2}{2 (T_i + \theta)}}}
     {\sqrt{\frac{m}{2 \pi T_i}}e^{-\frac{m_i (v_i^{\alpha+})^2}{2 T_i}}} 
\nonumber\\
&= 
\prod_{\alpha}^{n_i} \sqrt{\frac{T_i}{T_i + \theta}} e^{\frac{m_i(v_i^{\alpha+})^2}{2}\frac{\theta}{T_i (T_i + \theta)}}
\nonumber\\
&= 
1 + \sum_{\alpha}^{n_i} \frac{m_i (v_i^{\alpha+})^2 - T_i}{2 T_i^2} \theta + \caO(\theta^2)
\end{align}
The expansion in the last line allow us to keep the first order in $\theta$ and write the susceptibility as the unperturbed correlation
\begin{align}
\chi_\caO (t)
&= \left\langle \sum_{\alpha=1}^{n_i}\frac{ m_i (v_i^{\alpha+})^2 -T_i}{2 T_i^2} \caO(t)\right\rangle\nonumber\\
&\equiv\frac{1}{T_i} \left\langle \Gamma[\omega] \caO(t)\right\rangle
\,.
\label{eq:chi0}
\end{align}
where $\Gamma[\omega] \equiv  \frac 1 {2 T_i}\sum_{\alpha=1}^{n_i}[ m_i (v_i^{\alpha+})^2 -T_i]$.
We see that it is a correlation between the observable and a sum of kinetic-minus-bath temperatures.
In order to gain more physical insight, we may split the functional $\Gamma[\omega]$ in two terms with well-defined time parity,
\begin{align}
\Gamma[\omega] = \frac{S[\omega]  - K[\omega]} 2
\end{align}
with a time-antisymmetric component $S[\omega]$ and a time symmetric one, $K[\omega]$. This symmetries are understood by considering a time-reversed trajectory $\omega^R$, which corresponds to the evolution backward in time, with variables $\{x_i,-v_i\}$, from the final state of $\omega$.
By definition,
\begin{align}
S[\omega^R]&=-S[\omega]\,,\nonumber\\
K[\omega^R]&= K[\omega]\,.
\end{align}

We call $v_i^{\alpha-}$ the velocity of mass $i$ at the instant right before the update with index $\alpha$. In the time-reversed trajectory, that jump would thus be a transition from $- v_i^{\alpha+}$ to $-v_i^{\alpha-}$. 
It is easy to find that
\begin{align}
S[\omega] &=
\frac{1}{T_i}\sum_{\alpha} \left[\frac{m_i}2(v_i^{\alpha+})^2 - \frac{m_i}2(v_i^{\alpha-})^2\right]
\label{eq:S}\\
K[\omega] &= 
\frac{1}{T_i}\sum_{\alpha} \left[T_i - \left(\frac{(v_i^{\alpha+})^2 + (v_i^{\alpha-})^2}2\right)  \right]
\label{eq:K}
\end{align}
We see that $S[\omega]$ is a sum of kinetic energy gains (of the system) during interactions with the Andersen thermostat, all divided by temperature. This is the entropy production in the heat bath.
The interpretation of $K[\omega]$ is not as straightforward. We may call $K[\omega]$ the dynamical activity~\cite{lec05,mer05,gar09,ful13} or frenesy~\cite{bai10,bas15} of the systems: in our model it contains all deviations between the bath temperature and the kinetic temperature ``during'' a jump (i.e.,~the average of the two values before and after the jump). It can be seen as a measure of how agitated the system is in comparison to what it should be on average according to $T_i$.
With (\ref{eq:S}) and (\ref{eq:K}) one can reinterpret the susceptibility (\ref{eq:chi0}) as a sum of two correlations
\begin{align}
\label{eq:chi}
\chi_\caO (t) & = \frac{1}{2 T_i}\left[ 
\langle S[\omega] \caO(t)\rangle -
\langle K[\omega] \caO(t)\rangle 
\right]
\nonumber\\
& \equiv \mathcal{C^-}(t) + \mathcal{C^+}(t)
\end{align}
with 
\begin{align}
\mathcal{C^-}(t) &=\frac 1{2 T_i}\langle S[\omega] \caO(t)\rangle  \\
\mathcal{C^+}(t) &= - \frac 1{2 T_i}\langle K[\omega] \caO(t)\rangle 
\end{align}

In equilibrium at temperature $T_i=T$, for state observables $\caO(t) = \caO(\{x_i(t),v_i(t)\})$ (thus excluding functionals $\caO[\omega]$ such as integrated currents), we expect the combination $\frac 1 2 ( \mathcal{C^-} + \mathcal{C^+} )$ to turn into $\mathcal{C^-}$ 
(the susceptibility should be the time integral of a response function from Kubo's formula), namely
\begin{equation}
\chi_O(t) = \frac 1 T \langle S[\omega] \caO(t) \rangle_{\rm eq}
\end{equation}
which is the equilibrium correlation between observable and entropy produced into the environment (times the $1/T$ coming from the definition of thermal susceptibility).
Since 
\begin{align}
K[\omega] &= S[\omega] + \bar \Gamma[\omega]\nonumber\\
\textrm{with}\quad \bar \Gamma[\omega] & = \sum_\alpha [(v_i^{\alpha-})^2 - T_i]
\end{align}
to prove that our approach recovers the standard structure in equilibrium, we just need to show that $\langle \bar \Gamma[\omega] O(t)\rangle_{\rm eq} = 0$.
In equilibrium we can exploit the time-translational invariance of correlation functions as well as their invariance for time reversal. Using time reversal on $\bar \Gamma[\omega]$ changes its jump variables to those $(v_i^{\alpha+})$ after a jump. From these considerations we get
\begin{align}
\left\langle \sum_\alpha [(v_i^{\alpha-})^2 - T_i] \caO(t) \right\rangle_{\rm eq} & = \nonumber\\
\left\langle \sum_\alpha [(v_i^{\alpha+})^2 - T_i] \caO(0) \right\rangle_{\rm eq} & = 0
\end{align}
where the last term equals zero because velocities extracted by the Andersen procedure do not correlate with an observable at the beginning of the trajectory.

Before discussing some numerical results, we conclude this section by noting that the generalization to the perturbation of many degrees of freedom is trivial. If these have indices $i \in I$ and thermostats are independent stochastic processes,
\begin{align}
\chi_\caO (t)
&= \left\langle \sum_{i\in I}\sum_{\alpha_i=1}^{n_i}\frac{ m_i (v_i^{\alpha_i+})^2 -T_i}{2 T_i^2} \caO(t)\right\rangle
\,,
\label{eq:chi-many}
\end{align}
where now jumps have indices $\alpha_i$ that refer to their respective velocity index.

\color{black}
\section{Numerical results for the FPU model}
\label{sec:FPU}

As a toy model for heat conduction, we consider a FPU chain of coupled oscillators ordered from $i=1$ to $i=N$, each one with with mass $m=1$. Forces are determined by the FPU interparticle quartic potential
\begin{align}
U &= \sum^{N-1}_{i=1}U_{\rm FPU}(r_i)
\,, \nonumber\\
U_{\rm FPU}(r) &\equiv \left[ \frac{\kappa_2}{2}r^2+ \frac{\kappa_3}{3} r^3+\frac{\kappa_4}{4}r^4\right]
\,, \nonumber\\
r_i &\equiv x_{i+1} - x_{i} \,.
\label{eq:FPU-U}
\end{align}
The presence of $\kappa_3\ne 0$ yields an asymmetric potential and $\kappa_4 > 0$ makes it well concave for $r \to \infty$.
We selected $\kappa_2 = 27$, $\kappa_3 = -63$ and $\kappa_4 = 48$, and $N=10$ oscillators.
Thermostats are applied only at the two boundaries, with temperatures $T_1$ and $T_N$. These temperatures are in natural units, i.e.~we continue using $k_B = 1$.
We use the same value of the typical time $\tau=1$ between velocity renewals in both Andersen thermostats.
This time scale is of the same order of the oscillation times in the potential wells: indeed,
we use temperatures in the range $T=0.1\div 0.5$, corresponding to thermal velocities $v_T \approx \sqrt{T} \approx 0.3 \div 0.7$ and oscillation ranges $r_o \approx 0.4\div 1$ which are covered in timescales $\approx r_o/v_T$ of order $1$. 
Simulations are run with a much shorter time step $dt = 0.005$ and
data collection starts after a conservatively long relaxation that ensures having reached the stationary state. 
This is also guaranteed by the small size ($N=10$) of the FPU system. We thus should not be dealing with the issues of thermalization in one-dimensional long FPU systems~\cite{lep03,dha08,ben11a}.

In the following we show that the typical response of our FPU system takes place in time scales at least one order of magnitude longer than $\tau$. In fact, we observe a slow, scale-free convergence $\sim 1/t$ to the asymptotic $t \to \infty$ limit of the static response.
The thermal fluctuations introduced randomly by thermostats at the boundary velocities naturally propagate in the chain due to the FPU interaction between the oscillators.  A heat flux is set up within the chain if the two thermal reservoirs are at two different temperatures $T_1 \ne T_N$. 

%%%%%%%%%%%%%%%%%%%%%%%
\begin{figure}[!tb]
\centering
\includegraphics[width=8cm]{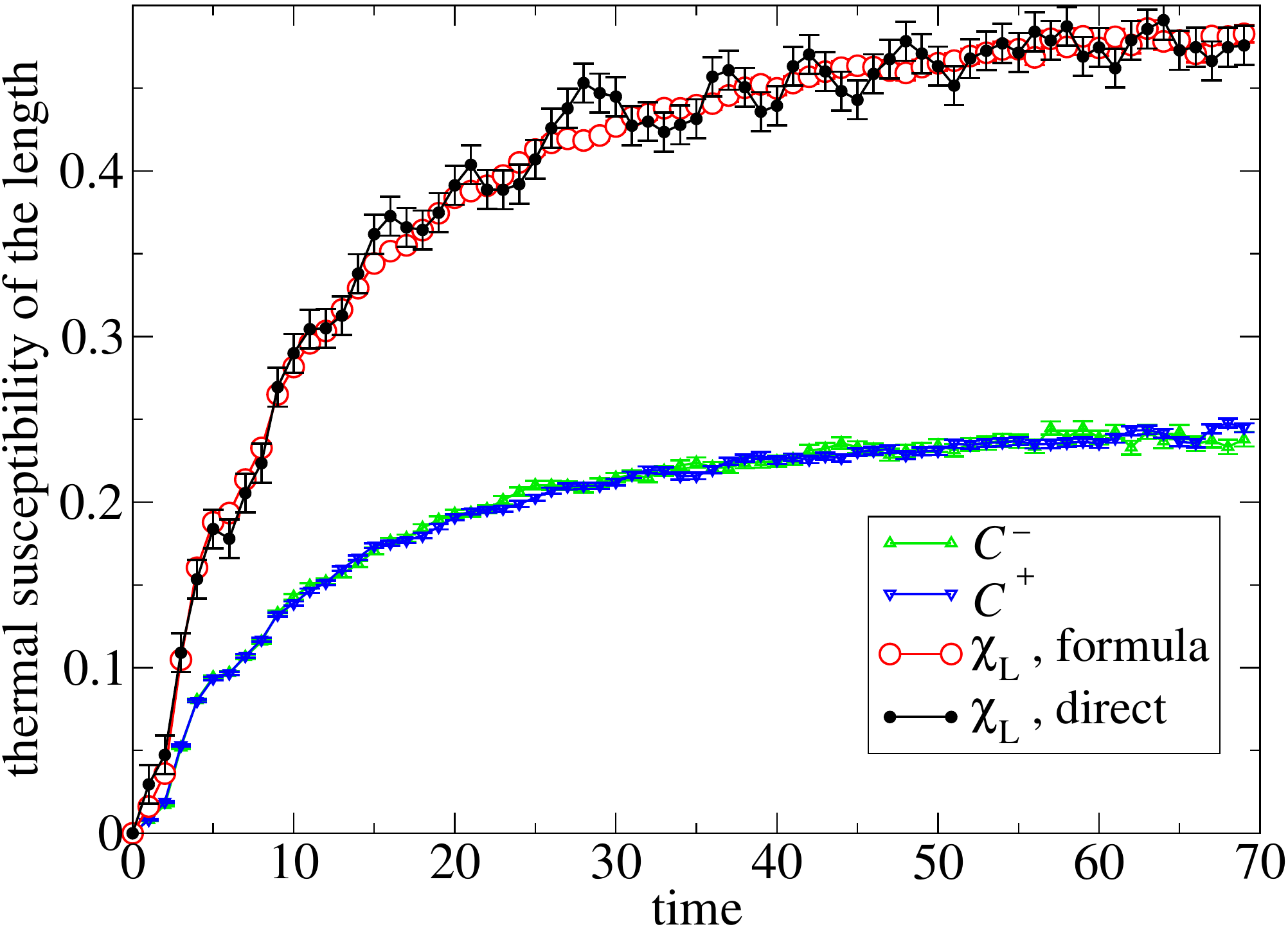}
\includegraphics[width=8cm]{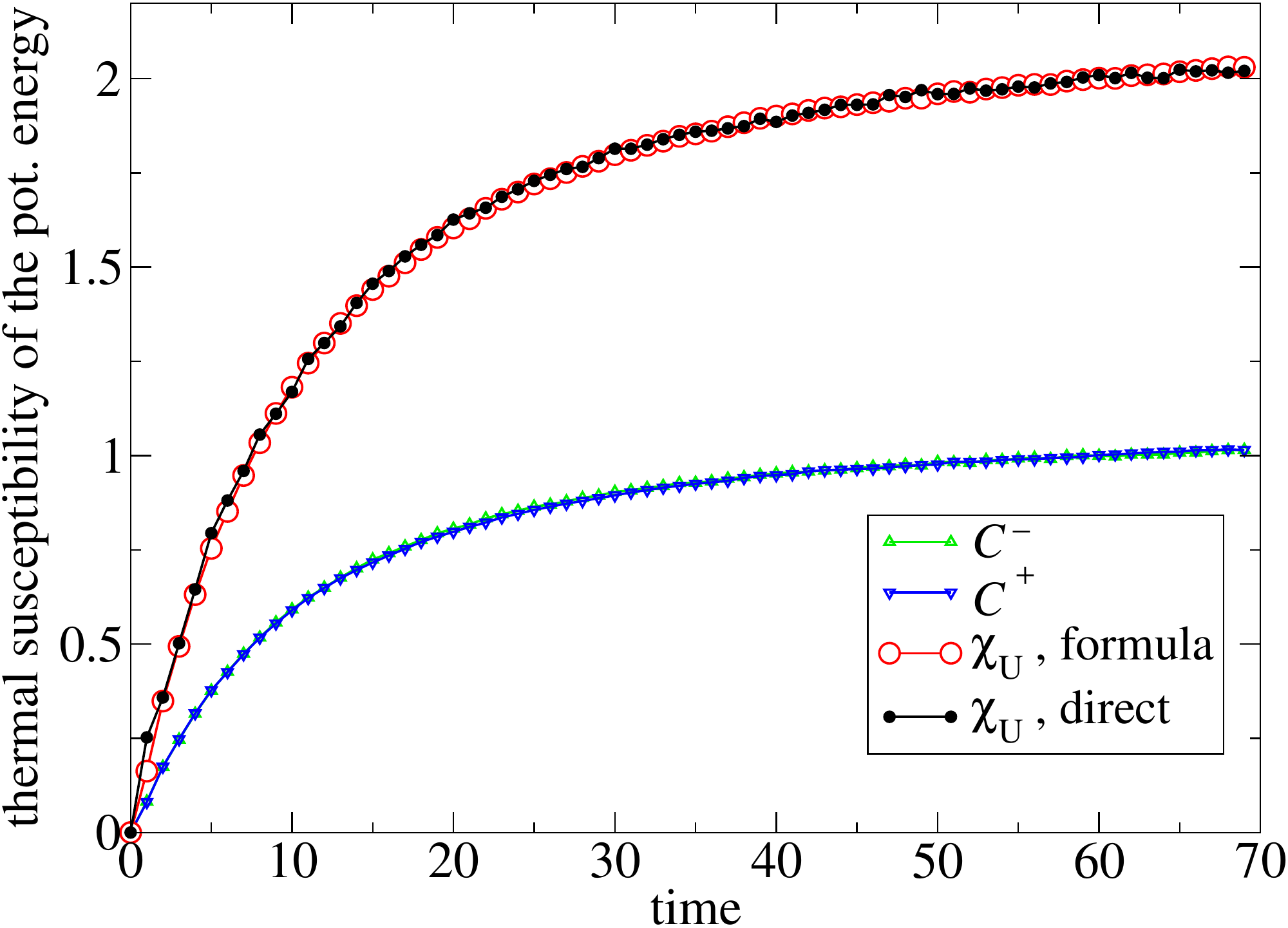}
\caption{Susceptibility at equilibrium ($T_1 = T_N = 0.1$) of the system length (top panel) and of the internal energy (bottom panel) to a variation of $T_N$. Black bullets represent the susceptibility computed by perturbing the system, while red circles represent the susceptibility estimated with our fluctuation-response formula.
Note the equality in equilibrium of the two term $ \mathcal{C^{-}}$ and $\mathcal{C^{+}}$ that compose the susceptibility.}
\label{fig:eqChi}
\end{figure}
%%%%%%%%%%%%%%%%%%%%%%%

%%%%%%%%%%%%%%%%%%%%%%%
\begin{figure}[!tb]
\centering
\includegraphics[width=8cm]{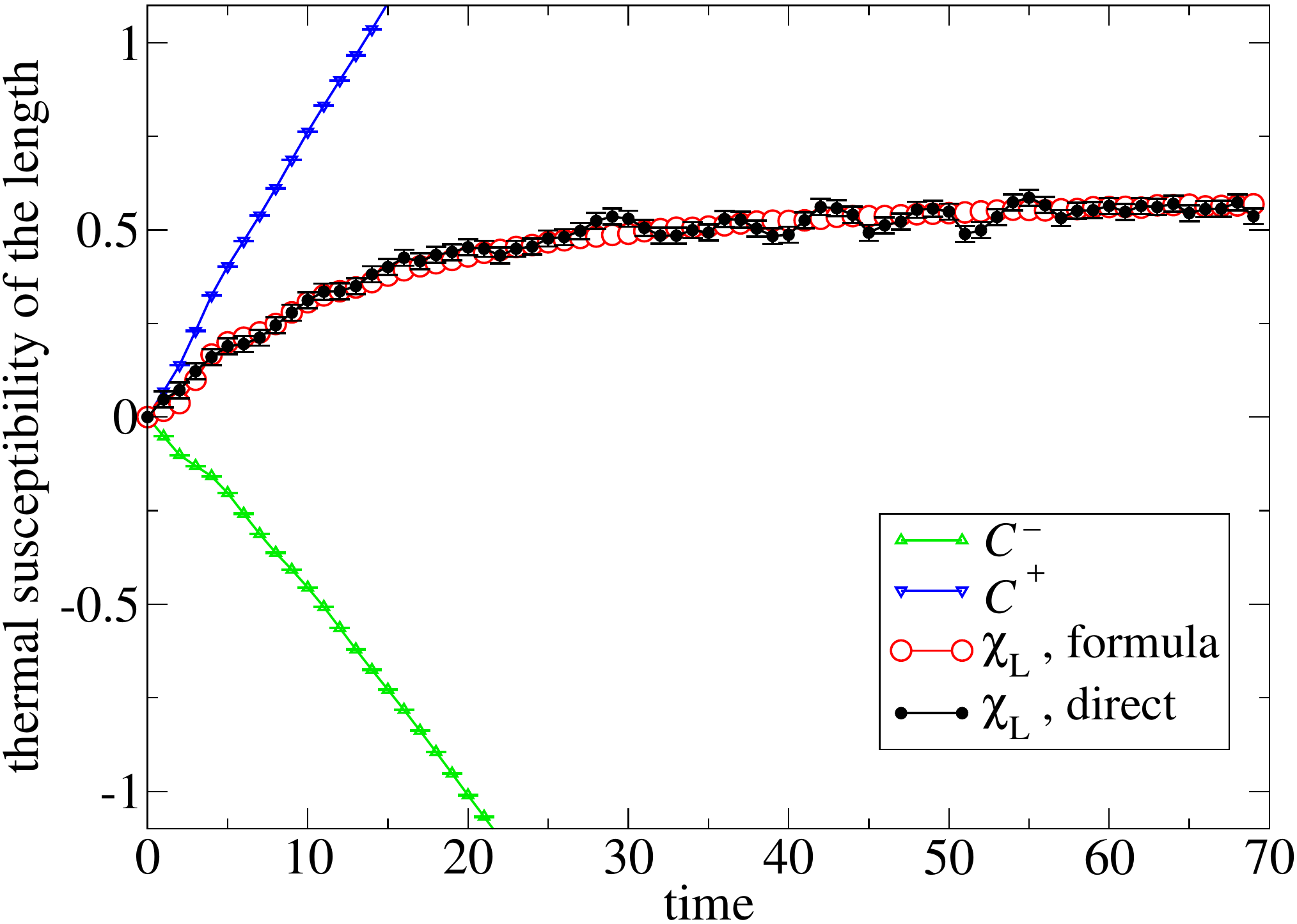}
\includegraphics[width=8cm]{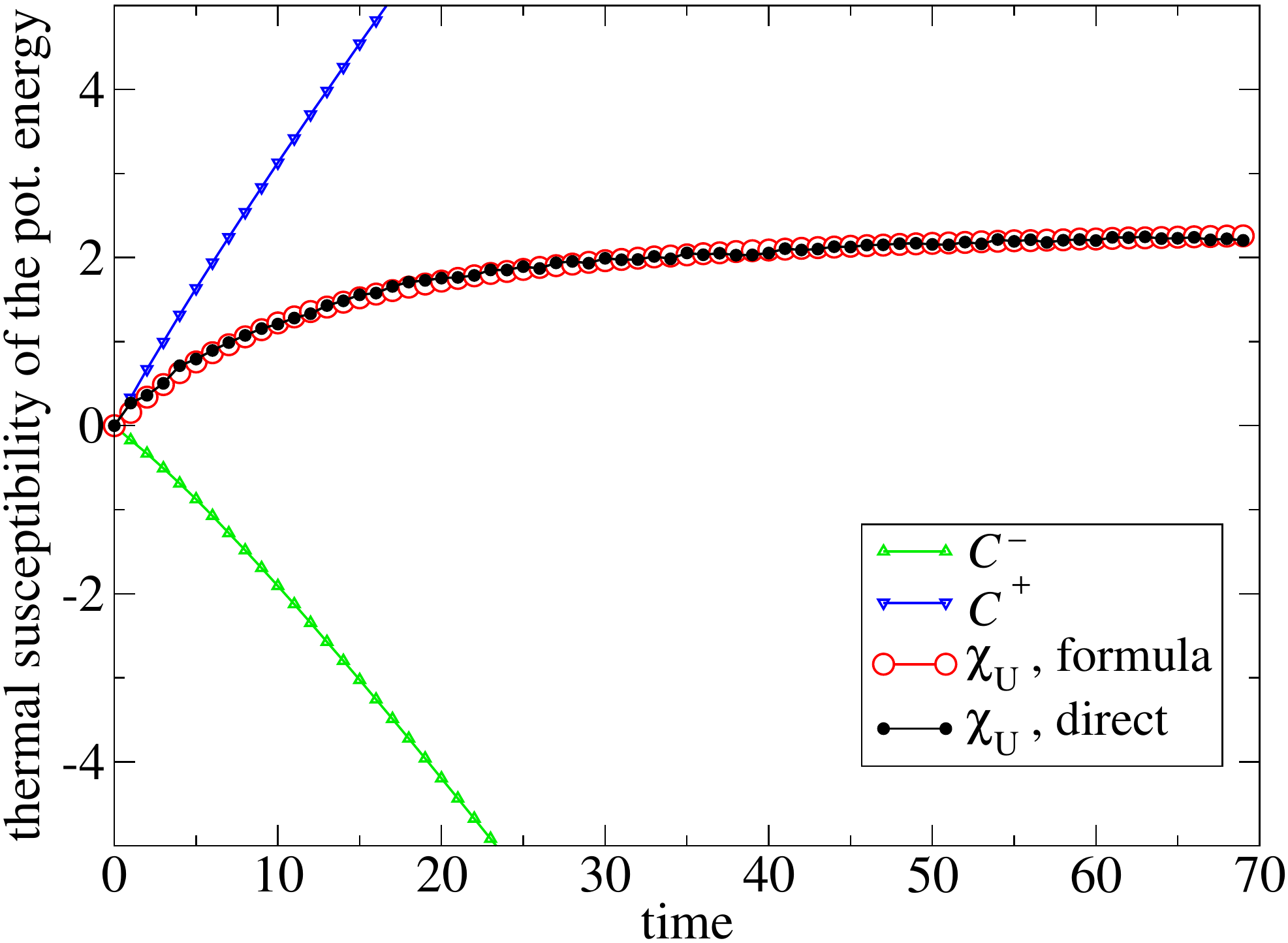}
\caption{Susceptibility far from equilibrium ($T_1 = 0.1$ and $T_N = 0.5$) of the system length (top panel) and of the internal energy (bottom panel) to a variation of $T_N$. Black bullets represent the susceptibility computed by perturbing the system, while red circles represent the susceptibility estimated with our fluctuation-response formula.}
\label{fig:neqChi}
\end{figure}
%%%%%%%%%%%%%%%%%%%%%%%

We perturb the temperature $T_N$, hence the perturbed degree of freedom in the notation of the previous section is $i=N$.
The two observables we consider are the chain length\footnote{Since $x_i$'s should be interpreted as displacements from equilibrium positions, the absolute length of the system is not specified and $L$ is more appropriately interpreted as the difference between the length and that at zero temperature.} $L\equiv\sum_{i=1}^{N-1} r_i =  x_N - x_1$ and the total potential energy $U$.
In the first case, the susceptibility $\chi_L(t)$ quantifies the thermal response of the system size, which is always an expansion because $\kappa_3<0$.  The susceptibility $\chi_U(t)$ of the energy instead generalizes the concept of heat capacitance.

To check the algorithm, we first look at equilibrium results. We set $T_1 = T_N = 0.1$ and, for the direct evaluation of susceptibilities, we use the temperature step $\theta=0.01$ for perturbations. We recover the correct equilibrium Boltzmann distributions (not shown) for kinetic and potential energies.

The susceptibilities computed without perturbing the system via equations (\ref{eq:S})-(\ref{eq:chi}) match those determined with a small perturbation, see Fig.\ref{fig:eqChi}. Moreover, at equilibrium, one notes that $\mathcal{C^{+}} = \mathcal{C^{-}}$ as expected. Therefore one may reconstruct the susceptibility from only the usual correlation
between the entropy production and the observable, i.e., the Kubo formula $\chi = 2 \mathcal{C^{-}}$.

Out of equilibrium, we do not find anymore $\mathcal{C^{+}} = \mathcal{C^{-}}$, rather the symmetric and antisymmetric terms may diverge dramatically, as shown for $T_1 = 0.1$ and $T_N = 0.5$ in Fig.~\ref{fig:neqChi}. Yet, with (\ref{eq:S})-(\ref{eq:chi}) we are able to correctly compute the susceptibility also in a strong nonequilibrium regime and without perturbing the system. One may also note that the susceptibility computed with the fluctuation-response formula, both in equilibrium and out of equilibrium, is smoother and affected by smaller errors than the susceptibility computed by perturbing the system.

%%%%%%%%%%%%%%%%%%%%%%%
\begin{figure}[!tb]
\centering
\includegraphics[width=8cm]{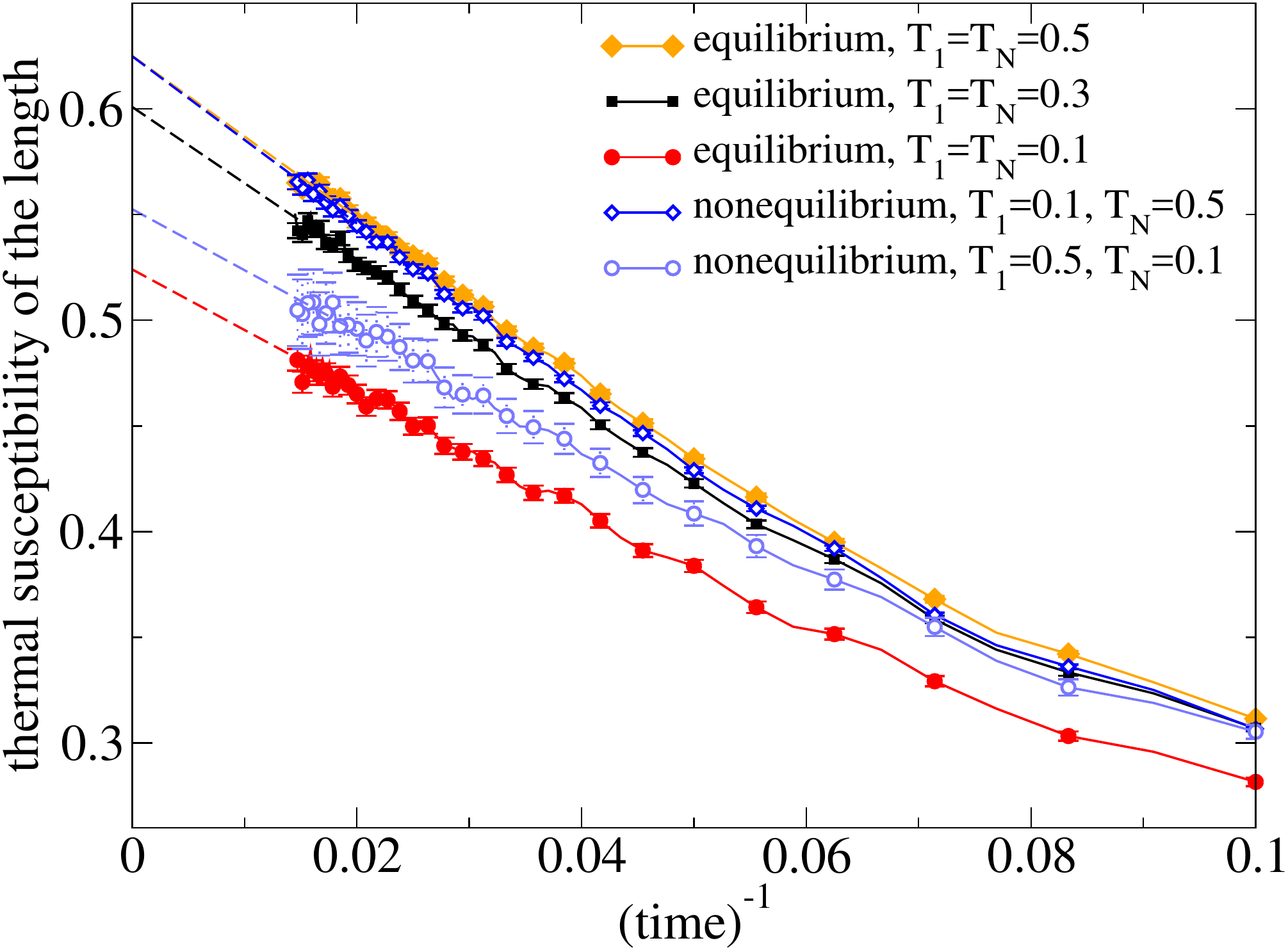}
\caption{Susceptibility of the system length to a variation of $T_N$ vs.~the inverse of time, for steady states at different boundary temperatures: three equilibrium regimes and two nonequilibrium ones (see legend). Dashed lines are linear extrapolations.}
\label{fig:chiL}
\end{figure}
%%%%%%%%%%%%%%%%%%%%%%%

Since the numerical results obtained with (\ref{eq:S})-(\ref{eq:chi}) are better than those collected via a real perturbation of the system, we use them for computing the susceptibility of the chain length to a variation of $T_N$ in different regimes. We then derive a static susceptibility
\begin{equation}
\chi_L^s \equiv \lim_{t\to\infty} \chi_L(t)
\end{equation}
The extrapolation to $t\to\infty$ is obtained by noting that data seem to converge toward the asymptotic value with a gap decaying $\sim 1/t$. A linear fit of the data as a function of $1/t$ is shown in Fig.~\ref{fig:chiL}. Having sampled at three different equilibrium conditions we note that $\chi_L^s$ is not constant in this model, rather it is increasing with $T=T_1 = T_N$: we get $\chi_L^s=0.524(3)$ for $T=0.1$, $\chi_L^s=0.601(2)$ for $T=0.3$, and $\chi_L^s=0.625(2)$ for $T=0.5$. 

The nonequilibrium cases suggest that $\chi_L^s$ is not very sensible to the temperature on the unperturbed side ($T_1$) if it is lower than $T_N$ (for $T_1=0.1$ and $T_N=0.5$ we get again $\chi_L^s=0.625(2)$). This seems not to be the case for the opposite condition $T_1=0.5$ and $T_N=0.1$, in which the perturbation of $T_N$ leads to a $\chi_L^s= 0.553(3)$ higher than that in equilibrium at $T_1=T_N=0.1$. This is a genuine nonequilibrium effect related to the heat flux coming from the side of $T_1>T_N$. Finally, note that both nonequilibrium cases are different from that in equilibrium at the average temperature $T=0.3$.

\section{Conclusions}
A thermal response relation can be easily derived for inertial systems driven by Andersen thermostats and it turns out to contain quite simple expressions. This contrasts with the difficulty of deriving a fluctuation-response relation for temperature variations in systems following a Langevin dynamics. A sound version of the latter (namely a theory devoid of finite time-step singularities) is currently available only for overdamped systems~\cite{fal16,fal16b,yol17}. 
Moreover, a linear response theory for systems with deterministic degrees of freedom coexisting with stochastic ones, as in the FPU model driven at the boundaries, seems impossible for Langevin systems~\cite{priv-com-GF}: by reguralizing path integrals as previously done~\cite{fal16,fal16b,yol17}, part of the perturbation is shifted to deterministic degrees of freedom, for which one cannot expand the path probability as normally done for stochastic degrees of freedom. 
This suggests that Andersen thermostats are an ideal tool for studying thermal susceptibilities, in particular in boundary-driven systems.

The example we analyzed, a FPU chain driven by two boundary temperatures, shows how to define the alike of heat capacitance and of thermal expansion susceptibility in a simple system carrying a heat flux. Numerical results suggest that the response to a change of a temperature is better estimated with our fluctuation-response relation than by actually perturbing the system (which is already not convenient in simulations, as it would also require more CPU time). Relaxing systems could be studied in the same framework.

\paragraph{Acknowledgements} We thank Gianmaria Falasco for useful discussions.

\vspace{0.5cm}
Both authors wrote this paper and contributed with analytical calculations and with analyses of numerical data. FDA implemented the simulations.

\bibliographystyle{spphys}       % APS-like style for physics
%%\bibliography{bib_noneq}

\end{document}